\pdfoutput=1
\documentclass[a4wide,twocolumn]{article}
\usepackage{amsmath}
\usepackage{graphicx}

\title{Energy-Efficient Data Transfers in Radio Astronomy with Software UDP RDMA\footnote{Preprint submitted to \textbf{Future Generation Computer Systems}}}

\author{Przemyslaw Lenkiewicz\\
  IBM Research - Netherlands \\
  \textit{lenkiewicz@nl.ibm.com} \and
  P. Chris Broekema \\
  ASTRON --\\
  Netherlands Institute for Radio Astronomy\\
  \textit{broekema@astron.nl}\and
  Bernard Metzler\\
  IBM Research - Zurich\\
  \textit{bmt@zurich.ibm.com}}

\begin{document}
\maketitle

\begin{abstract}
Modern radio astronomy relies on very large amounts of data that need to be  transferred between various parts of astronomical instruments, over distances that are often in the range of tens or hundreds of kilometres. 
The Square Kilometre Array (SKA) will be the world's largest radio telescope, data  rates between its components will exceed Terabits per second.
This will impose a huge challenge on its data transport system, especially with regard to power consumption.
High-speed data transfers using modern off-the-shelf hardware may impose a significant load on the receiving system with respect to CPU and DRAM usage.
The SKA has a strict energy budget which demands a new, custom-designed data transport solution.
In this paper we present SoftiWARP UDP, an unreliable datagram-based Remote Direct Memory Access (RDMA) protocol, which can significantly increase the energy-efficiency of high-speed data transfers for radio astronomy.
We have implemented a fully functional software prototype of such a protocol, supporting RDMA Read and Write operations and zero-copy capabilities.
We present measurements of power consumption and achieved bandwidth and investigate the behaviour of all examined protocols when subjected to packet loss. 
\end{abstract}

\section{Introduction}
\label{sec:intro}
Modern radio telescopes, such as the LOw Frequency ARray (LOFAR)~\cite{vanHaarlem:2013dsa} and the upcoming Square Kilometre Array (SKA)\cite{Dewdney:2009} are in essence large-scale distributed sensor networks, characterised by large numbers of receivers producing staggering amounts of data.
This data is often generated by custom hardware in remote areas, while processing it into usable science data is done in data centres in nearby cities.
Receiving and processing these streams is a computationally intensive task that may consume considerable amounts of energy.
A current state-of-the-art example is LOFAR: 54 antenna stations produce around 250\,Gb/s of sensor data in total, to be transported over 65\,km to the central processor.
The Square Kilometre Array (SKA) will produce much more data, to be transported over much longer distances.
This data stream, about 3\,Tb/s per telescope, is to be transported from the Western Australian desert to Perth, and from the Karoo desert to Cape Town, both several hundred kilometres away.

Whereas the available compute capacity in current telescopes is often limited by available capital, in the SKA it will likely be limited by available energy.
Experience with the LOFAR radio telescope has shown that receiving large volumes of data may consume significant compute resources~\cite{Romein:10,Romein:11a}.
These consumed resources cannot contribute directly to the science result.
It is therefore useful to investigate ways to minimize the resources, and energy, required to receive streaming radio astronomical data.
Reducing the data transfer protocol overhead allows more of the precious resources to be dedicated to scientific processing.
More energy-efficient handling of incoming data can be directly translated into additional science output within the limited available energy budget.

Typical implementations of the network protocol stack such as the Linux\text{*} IP stack are designed with robustness and security in mind.
Strict separation between user and system resources is maintained.
Received data is copied several times and will trigger several interrupts and context switches before the user application gains access to it.
In the IBM\text{*} Blue Gene/P supercomputer a different bottleneck, namely software handling of Translation Lookaside Buffer (TLB) misses, was mitigated by bypassing conventional kernel processing~\cite{Kazutomo:10}.
This was used to significantly decrease compute resources consumed while receiving LOFAR data.
In this paper we propose a similar approach aimed at the Square Kilometre Array.
The Linux IP stack requires significant resources, in particular while receiving large volumes of sensor data.
We propose to bypass the host operating system and place data directly into user memory.
While this also bypasses several of the security features that are essential in a typical network stack, in a tightly controlled and private network, such as found in a scientific instrument, these are less crucial.
We expect a reduction in resource consumption and therefore a reduction of consumed energy.
Since the SKA Science Data Processor is expected to be bound by very tight budgets, in particular in available energy, reducing the computational cost of receiving data would allow for more science, improving the scientific efficiency of the instrument.

Remote Direct Memory Access (RDMA) technology has been essential in high-performance networking to resolve similar issues, namely to allow higher bandwidth, lower latencies and lower CPU utilization.
RDMA-capable network interface controllers (RNICs) provide this by moving data directly from the user space memory of one machine to that of another, without involving either of the host operating systems. 
The application layer is involved only on the side where the request is issued and it can access the contents of memory buffers on a different host thanks to memory pre-registration.
The RDMA technology is a very good example of how the data movement process can be optimized for a specific scenario, helping to utilize the full capabilities of the hardware.
However, the currently-available RDMA solutions lack some of the features that are necessary for a scenario such as the SKA. 
In particular, the target scenario requires a more efficient handling of the expected very high bandwidth-delay product of the data transfer channel, and imposes application specific requirements on time sensitive, partial data transfer reliability~\cite{Broekema:2012:EHP:2286976.2286982}.

In this paper we address the data transport challenges for modern radio astronomy instruments.
We introduce a possible solution that measurably reduces the consumption of CPU resources and energy associated with that data transport.
In particular, we design and implement an efficient communication protocol for transferring high rates of astronomical data over long distances with the goal of being more energy-efficient at the receiving end.

The main contributions of this paper are:
\begin{enumerate}
  \item we design and prototype in software a partially reliable, RDMA-based transport protocol suitable for modern radio astronomy applications;
  \item we present experiments with results showing that the energy-efficiency of the prototyped transport stack is improved compared to standard UDP data transfer;
  \item we argue that further, more dramatic improvements in efficiency are possible when support for this protocol is implemented in hardware.
\end{enumerate}

\section{The Square Kilometre Array}
\label{sec:TheSKA}

The Square Kilometre Array (SKA) is a new-generation radio telescope which is currently being designed by an international science and engineering team. 
Construction is expected to commence in 2019 and the first phase is expected to become operational from 2022.
SKA phase 1 (SKA1) will consist of two instruments: SKA1-Low, located in Western Australia and SKA1-Mid in South Africa~\cite{ChrisSDP}.  
SKA1-Low is an aperture array instrument consisting of 512 stations, each with 256 dual-polarised antennas, operating between 50 and 350\,MHz.
The stations will be 35\,m in diameter and placed, at most, 65\,km apart.  
The antenna signals are coherently summed per station into a station beam.
SKA1-Mid will consist of 133 dishes, with an additional 64 MeerKAT~\cite{5109671} dishes to be integrated into the SKA1 instrument, each with a diameter of 15\,m (13.5\,m for the MeerKAT dishes), capable of receiving signals between 350\,MHz and 14\,GHz.
The processing of data produced in SKA1-Low and SKA1-Mid is similar and therefore we will not distinguish between them in this paper.

\begin{figure}
\centering
\includegraphics[width=3.2in]{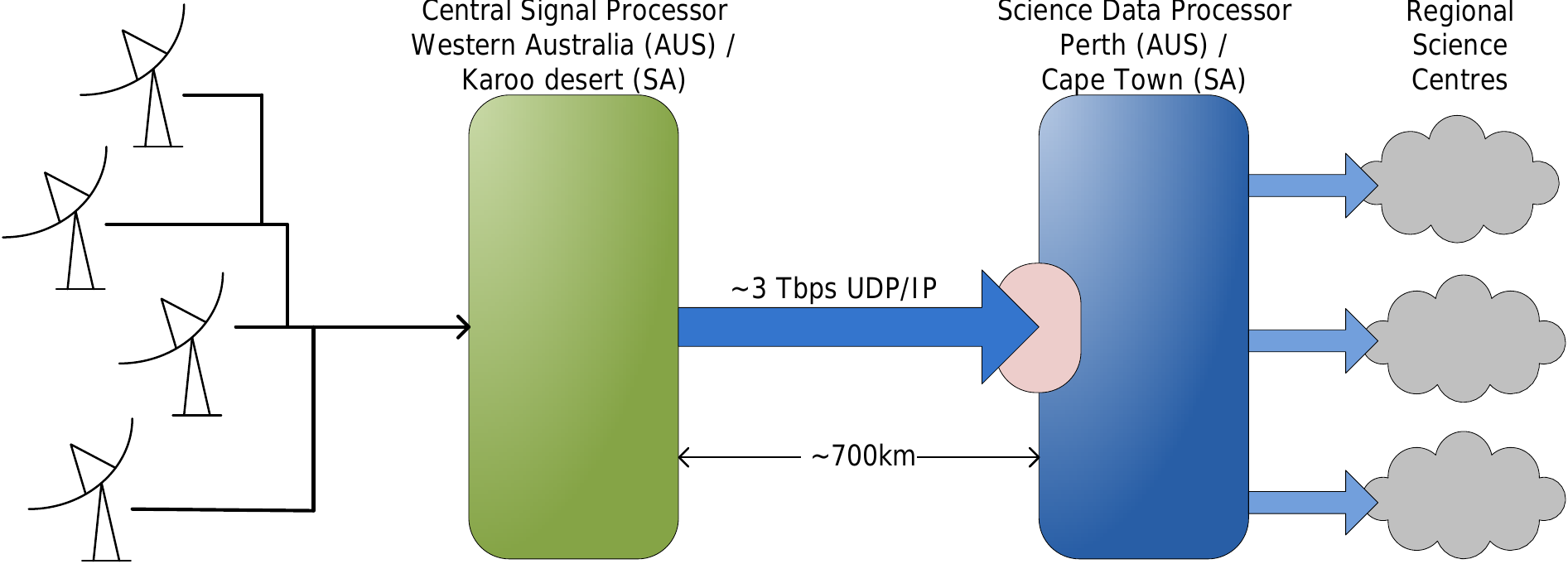}
\caption{Simplified data transport chart in the SKA, leading from the SKA stations to the Central Signal Processor and the Science Data Processor.}
\label{fig:SKADataFlow}
\end{figure}

A simplified diagram depicting the SKA data flow is presented in Fig.~\ref{fig:SKADataFlow}.
Data from the receivers is transported to the Central Signal Processor (CSP), located in a radio frequency-shielded building at the centre of the telescope. 
Here, data from all receiver or station pairs are combined into visibilities by the CSP correlator.
The resulting data is transported to the Science Data Processor (SDP), one for each instrument, located in Perth and Cape Town, several hundreds of kilometres away.
Each of the SDP instantiations receives a continuous data stream of about 3\,Tb per second.

The SDP produces science-ready calibrated data products for analysis by the radio astronomer, a task that is highly data intensive and is expected to require compute resources in the 100 PetaFlop range.
Data from the SDP are distributed to Regional Science Centres for further analysis and dissemination. 

Whereas the central processors in current telescopes are generally limited by available capital budget, it is expected that the SKA SDP will be limited by available energy.
Considering this strict budget, improving the energy-efficiency of secondary tasks not directly related to the science output can have significant indirect impact on the scientific efficiency of the SKA SDP.
In Section \ref{sec:intro} we cite previous work that showed that receiving large volumes of instrument data requires significant compute resources.
These consumed resources, and this fraction of the energy budget, do  not directly contribute to the scientific output of the SDP.
By avoiding a well known system bottleneck: the Linux IP stack and the associated kernel overhead, we reduce the energy  required to receive streaming data typical in radio astronomy.
As a direct consequence, more of the limited energy budget is available for scientific computing, leading to increased scientific output of the SKA SDP within the energy budget.

Considering the large distances and volumes of data, it is not feasible to use a reliable data transport protocol for the data transport between CSP and SDP.
This would require constant buffering of the transmitted packets at the sending side until confirmations from the receiving side arrive.
In a highly optimized real-time environment, such as the CSP correlator system, this would incur very significant cost and performance overheads.
The chosen transmission protocol for this data stream is therefore unreliable, based on UDP/IP over Ethernet, with far lower sender-side overheads.
At this point the transported data is highly redundant.
Loss of a fraction of this data will result in reduced signal-to-noise ratio in the end-product, but this is, within reason, acceptable.

The specific requirements for this particular SKA data transport component can be summarised as follows:
\begin{itemize}
\item very high data rates, several Tb/s
\item strictly uni-directional traffic
\item prioritizing bandwidth over latency 
\item desire for very high energy-efficiency
\item full reliability is not crucial, some data loss is tolerable
\end{itemize}

In the remainder of this paper we investigate how an existing industry standard RDMA implementation can be modified in such a way that it can be used to transport SKA specific data streams.
By avoiding a known bottleneck we expect to save a measurable amount of computational resources and energy.
This can immediately be translated into increased scientific performance for the same investment.

\section{RDMA, iWARP and SoftiWARP}
\hyphenation{Soft-iWARP}
Receiving multiple high-bandwidth UDP/IP data streams requires significant CPU resources. 
Since CPU cycles can be translated into consumed energy, it can be assumed that a more efficient way to receive large data streams will consume less energy. 
In addition, compute resources spent on receiving data cannot be
utilised for data reduction or processing.

Implemented as an operating system service, the Linux network I/O stack
was designed with the main focus on robustness and security while maintaining good
performance. Applications access network services via the socket API.
To achieve separation and protection, all communication data is copied
between application buffers (user space memory) and operating system
(kernel) memory within the socket layer. 
On the transmission path, after copying data into the kernel, network
protocol output processing packetises the data, stores it for potential retransmission and informs the network adapter to fetch the packets for wire transmission.
In the network packet input path, data are first moved from the network
card into kernel memory and an interrupt is issued, which handles network protocol processing within the kernel. 
As a result of protocol processing, kernel data buffers containing
the received data are queued to the socket receive queue for
application retrieval.  Within a system call, the application eventually
copies those data from kernel memory to application receive buffers, which typically involves waking up the application thread waiting for data reception.
Both in the sending and receiving path, traversing the Linux networking stack incurs non-negligible overheads (interrupt handling, context switches, network protocol processing, data copy operations), which degrades application-available CPU processing power, while limiting achievable communication bandwidth and adding to end-to-end communication latency.
Moving the data directly between the network device and application buffer would avoid such overheads, but if not done properly, would violate the data protection and separation principles of the operating system.
However, in a tightly controlled and private environment, such as in
a scientific instrument, these limitations might be acceptable.

In the past decade the Remote Direct Memory Access (RDMA) technology has
been gaining more and more relevance in the field of high-speed communication.
Its development was driven by the need for high throughput and low latency networking, especially in High Performance Computing.
RDMA provides this by moving data directly from the user space
memory of one machine to that of another, without involving the host operating system and minimising host CPU usage. 
The application layer registers memory buffers with the local RDMA-capable network interface controller (RNIC) for remote write or read access. 
Under the control of local and remote RNIC, RDMA write operations transfer data from a local buffer to a tagged remote buffer that was previously signaled by the peer, whereas the RDMA read operation transfers data from a tagged remote buffer to a tagged local buffer.
The application layer is involved only on the side where the request is issued.
Any application buffer used as a source or target for an RDMA operation must be pre-registered with the local RNIC device, and is typically pinned into physical host memory. This allows the RDMA device to access the buffer in physical memory without further OS intervention.
To allow overlapping communication and computation, RDMA offers
an asynchronous communication interface. RDMA operations are posted
as Work Requests (WRs) to a communication endpoint and are asynchronously processed by the RDMA device. Work completions are signalled and retrieved asynchronously as well.

RDMA is provided through several network technologies, including
Myrinet~\cite{myrinetboden1995}, Infiniband~\cite{infinibandpfister2001introduction}, 
RDMA over Converged Ethernet (RoCE)~\cite{rocesubramoni2009rdma,rocebeck2011performance} and iWARP~\cite{iwarpgross1998,iwarprashti200710}. The functionality and performance of these standards has been evaluated and compared in various studies~\cite{comparendliu2003performance,comparerashti200710}. Well-known programming interfaces, like the Message Passing Interface, may be used in order to access the RDMA functionality on different hardware~\cite{mpiliu2004high}.

Both RoCE and iWARP are deployed over Ethernet, which makes them
very interesting candidates for the SKA data transport service.
RoCE defines the transmission of InfiniBand packets directly
over Ethernet, which limits its scope to the Ethernet
broadcast domain and thus leaves it non-routable. To solve that
issue, a recent protocol extension (RoCEv2) puts it on top of the UDP/IP protocols.
On the other hand, iWARP defines RDMA operations on top of TCP/IP
networks, giving it the advantage of being compatible
with the existing Internet infrastructure. Unfortunately, both
RoCE and iWARP rely on the implementation of a rather
complex protocol state machine (TCP or InfiniBand) meant to provide
a level of data transmission reliability which is not needed and even
obstructive for the intended use: data to be transmitted have a 
limited relevance in time -- in case of partial data loss
the protocol should favour the transmission of new data over
the retransmission of lost fragments. Lost data fragments shall result
in just dropping the entire affected application level message at
RDMA protocol level, while keeping the end-to-end connection intact.

In our work towards an energy-efficient protocol for modern radio astronomy we have chosen the iWARP standard as the baseline, but extended it with an unreliable service. 
This was achieved by replacing the TCP protocol with UDP and modifying the semantics of the RDMA application interface.

\subsection{Implementation of iWARP in software}
Although the full range of advantages of RDMA is only available through
hardware support for iWARP (in order to offload I/O and protocol processing 
from the CPU), a software implementation can also be beneficial.
iWARP is still a relatively young technology and therefore it is useful
 to be able to rely on a software solution for testing and development
purposes. Furthermore, the software version can be introduced in the
less stressed parts of the infrastructure, whereas the more utilised
parts would be equipped with iWARP-capable NICs -- provided that the
software implementation can operate in such a mixed scenario.
Thanks to the RDMA semantics and the asynchronous API, even a software
implementation can provide benefits such as a zero-copy data transmit path and less application interaction/scheduling, which can lead to increased performance and lowered CPU load and power consumption.
Software iWARP can also be used for migrating existing applications
to the RDMA interface without the need for RDMA hardware. Finally, it can ease the development of new, experimental extensions to the RDMA stack without hardware prototyping.
The SKA scenario is a good example of such a case, as we want to experiment with an implementation of iWARP that is tailored specifically for our needs. 

The idea to implement the iWARP protocol fully in software has already been explored and there are solutions available, such as the Software iWARP implementation by the Ohio Supercomputing centre~\cite{OhioiWarpinSoftware},\cite{OhioiWarpinSoftware2} or the SoftiWARP (SIW)~\cite{BernardSoftRDMA} implementation by IBM Research.

It is important to note that a software implementation of the iWARP protocol will most likely not guarantee a power efficiency to meet the energy budget requirements of the SKA. However, this choice is sufficient for experiments on the points relevant for the scientific instrument, namely CPU utilization, power consumption and behaviour under packet loss. The final solution, one that can be incorporated in the design of SKA, should rely on hardware support. This work should of course be seen as a step towards such a solution, as all the created code will be made available in a public repository.
\subsection{SoftiWARP}
The work presented in this paper is based on the SoftiWARP open source
software implementation of the iWARP protocol suite, developed at
IBM Research - Zurich and available from GitHub\footnote{https://github.com/zrlio/softiwarp}.
SoftiWARP comprises two main building blocks: a kernel module, which
implements the iWARP protocols on top of TCP kernel sockets, and a user
level library. 
SoftiWARP integrates with the industry standard OpenFabrics\footnote{https://www.openfabrics.org} RDMA host stack and thus exports the OpenFabrics RDMA API to both user space
and kernel space applications.
Due to close integration with the Linux kernel socket layer, SoftiWARP
allows for efficient data transfer operations. On the sending side,
it supports zero copy data transfers out of application buffers.
On the receiving side, the implementation makes use of target buffer
address information available with the RDMA protocol headers: the packet
payload is directly copied from their in-kernel representation ({\tt sk\_buff}) to the final application buffer without scheduling the receiving application.
Since the implementation conforms to the iWARP protocol specification,
it is wire compatible with any peer network adapter (RNIC) implementing
iWARP in hardware.
\subsection{Implementing an unreliable connected SoftiWARP service}	
In order to fulfil the requirements of the SKA we have defined and implemented a new unreliable, connection oriented RDMA transport protocol based on SoftiWARP.
Here, communication between hosts is implemented over UDP kernel sockets instead of the reliable, connection-oriented TCP.
The differences in using TCP and UDP for long-distance data transfers is a well-studied and documented matter. As we have described before, the reliability characteristic of TCP data transfers can result in a set of undesirable features. Known problems includee.g. a poor utilization of network capacity when using TCP in long-distance transfers~\cite{TCPUDP}, the need to use multiple streams in high-speed, long-distance network paths~\cite{TCPUDP2}, or the difficulty of management of the used buffer size to assure optimized delivery~\cite{Francini20123076}. To overcome some of these issues, the use of UDP in long-distance, high-speed data transfers has been proposed~\cite{Gu20071777,1137760}. 

In the work described in this article we are implementing a first connection
oriented RDMA protocol which incorporates an unreliable data transport layer.
We are able to perform RDMA Write and Read operations over this unreliable
transport service.
In SoftiWARP UDP the unreliable connection is used both for the connection
management operations, as well as the data transfer.
After connection setup, the application data transfer does not enforce
reliability, but is implemented in an unreliable, message-oriented manner: 
the sender segments the RDMA message into a set of UDP datagrams, which are 
reassembled on the receiver side into the original message and, if completely received, 
delivered to the application. Messages which remain incomplete due to UDP packet 
loss are silently dropped at the receiver.

To retain the efficiency of the original implementation, any inbound, in-sequence 
data is still directly placed into the application target buffer without 
intermediate queueing.
At API level, error handling has been implemented as simply as possible:
if a message remains incomplete due to data loss or corruption, the content 
of the target buffer remains undefined. 
If the lost message belongs to an RDMA Send/Receive operation, the current Receive
operation remains incomplete and the receive buffer gets re-used for placing the 
next inbound RDMA Send. 
Corrupted RDMA Write messages just leave the application buffer in an undefined state. 
While originally not defined for the iWARP protocol,
an 'RDMA Write with Immediate Data' operation might further improve the
handling of unreliable RDMA Writes at the target side: only if the RDMA
Write operation completes successfully, is the 'Immediate Data' delivered
to the application indicating the complete placement of a new RDMA Write.
These data could carry additional application level information such as
a message sequence number. Only InfiniBand and ROCE currently define
these optional 'Immediate Data' semantics for RDMA Writes. 
Therefore, it is currently up to the application to detect corrupted data
placed via RDMA Writes.

Unreliable RDMA Read operations are currently supported at an experimental level only. First of all, this operation is not required for the SKA use case: data streaming
is strictly uni-directional and only dictated by the sender delivering
radio-astronomic data to a data processing entity. 
Secondly, supporting unreliable RDMA Reads requires a further extension of the protocol state machine at the RDMA Read initiator side, since it must detect permanently lost RDMA Read Request/Response pairs. 
A timer based detection of message loss appears to be a viable solution to the problem, but is currently not implemented.

The extended SoftiWARP implementation runs on both UDP and TCP and
allows the selection of reliable connection (RC) or unreliable connection (UC)
services on a per connection basis. 
For the UC service, the client side must first create a connection endpoint with an appropriate OpenFabrics service attribute, namely {\tt IBV\_QPT\_UC}, which represents an Unreliable Connection Queue Pair.
On the server side a listener endpoint for the same service type must
exist. If the client connects its endpoint with the listener,
a new server side endpoint will result, which is associated with the
connecting client endpoint. 
After connection setup, both sides can use the new RDMA association for unreliable data transfer operations.

\section{Experiments}
\label{sec:experiments}
In this section we present in-depth tests of SoftiWARP UDP and analyse how a software implementation of the iWARP standard is able to perform in terms of achieved bandwidth and power consumption in comparison to standard TCP and UDP sockets. Our test platform comprises two servers equipped with Intel\text{*} Xeon E3-1240 v3 CPUs running at 3.40\,GHz, 16\,GB RAM and Chelsio T5-580 40\,Gb RDMA-capable Ethernet cards. The machines are interconnected with a direct connection using a QSFP+ cable. The tests have been performed with: 
\begin{itemize}
\item The Netperf\footnote{http://www.netperf.org} benchmark tool with additional tests implemented which carry traffic over RDMA protocols, both over TCP and UDP,
\item The LOFAR telescope traffic generator, which creates data packets at rates that correspond to that of a LOFAR telescope station. TCP and UDP sockets as well as TCP and UDP iWARP are supported for data transport.
\end{itemize}
We use two measurement points in our experiments to precisely assess the energy consumption of the data transfers. Using the RAPL Technology~\cite{RAPLIntel} the values from the Intel processor's registers can be read and the power consumption of the CPU and DRAM can be estimated in a very accurate way. We use the Performance Application Programming Interface (PAPI) library\footnote{http://icl.cs.utk.edu/papi/} and the Likwid tool\footnote{https://github.com/RRZE-HPC/likwid} to read the power meters. We have also constructed a custom-made power meter based on an Arduino board and voltage sensors attached to the PCI-Express slot. Using this device we can measure the power consumption of the NIC with an accuracy of 1/100\,Watt and 1\,millisecond sampling rate.
\subsection{Power consumption of the Chelsio T5}
\begin{figure}[h]
\centering
\includegraphics[width=0.95\columnwidth]{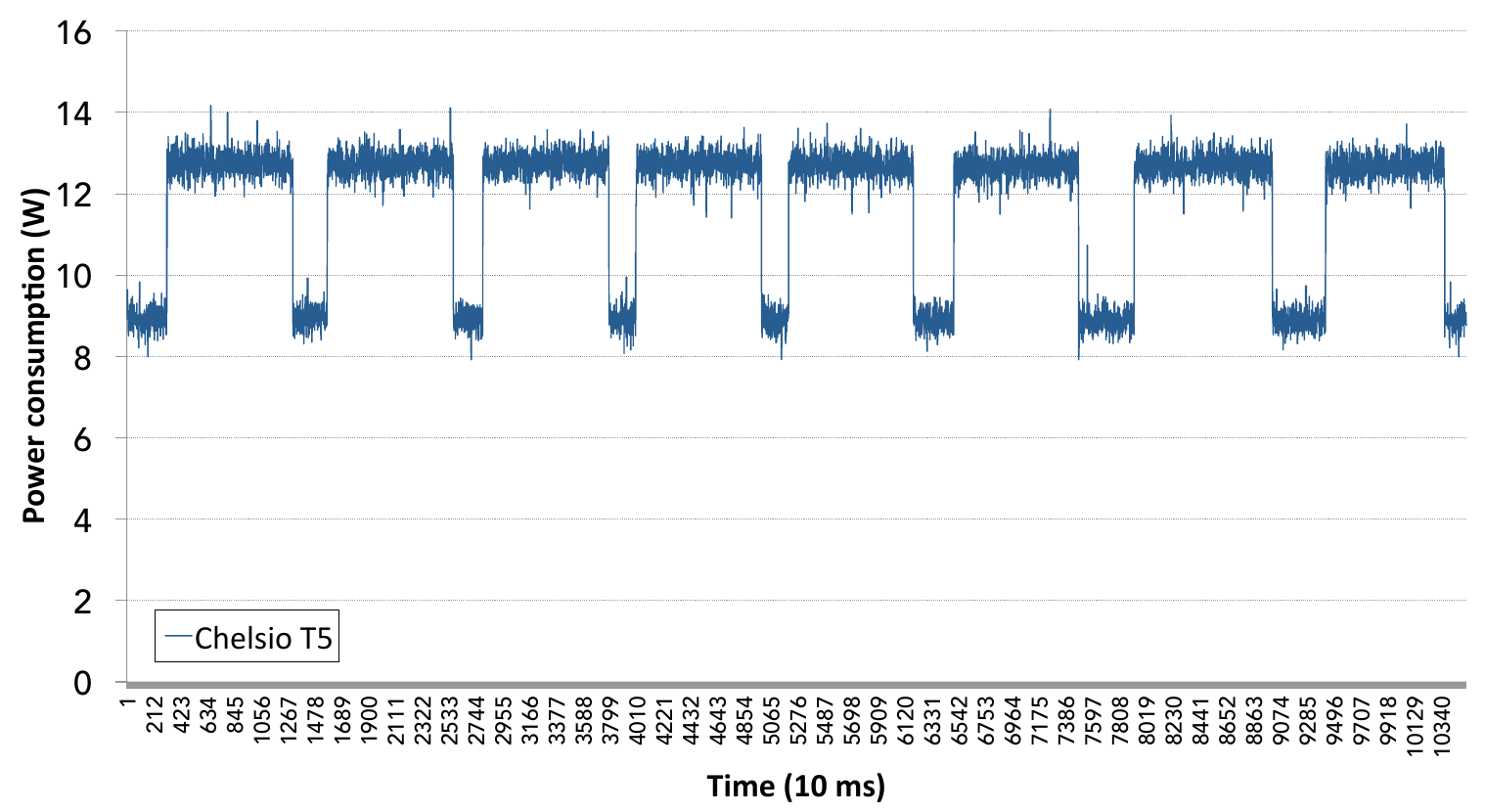}
\caption{Power consumption of Chelsio T5 during eight consecutive tests
  using Netperf (tests 1-6) and the LOFAR traffic generator
  (tests 7-8).}
\label{fig_T5bidirectional}
\end{figure}

The power consumption of the Chelsio T5 NIC has been measured using the power meter mentioned in the previous section, under numerous different test scenarios. The results of these tests are shown in Fig.~\ref{fig_T5bidirectional} in a consecutive manner. The blue line presents the trace of power consumption of the Chelsio T5 NIC. The value of 9\,W shows the idle state of the NIC and each peak of around 13.5\,W represents one test being carried out. 
Peaks  1 to 6 represent Netperf tests over different transport protocols in the following order:
SoftiWARP TCP, sending side; SoftiWARP UDP, receiving side; TCP sockets, sending side; TCP
sockets, receiving side; Hardware iWARP, sending side;  Hardware iWARP, receiving side. Tests 7 and 8 represent 50 instances of
the LOFAR traffic generator, first the sending side, then the receiving side. 

We can see from Fig.~\ref{fig_T5bidirectional} that the power consumption of the NIC card is very similar in all cases and doesn't depend on the kind of transport protocol used. 
Further tests have been performed with varying message sizes and all available transport methods, on both the sending and receiving side. 
All of them have shown nearly identical results of 9\,W for idle state and 13.5\,W for full link speed. Therefore, we can conclude that the power consumption of the RNIC is very consistent and doesn't show a dependency on the type of traffic. In the following sections we will focus only on the CPU and DRAM power consumption, as this is where all of the tested protocols show significant differences. 
\subsection{Radio astronomy data flow}
\begin{figure}[h]
\centering
\includegraphics[width=\columnwidth]{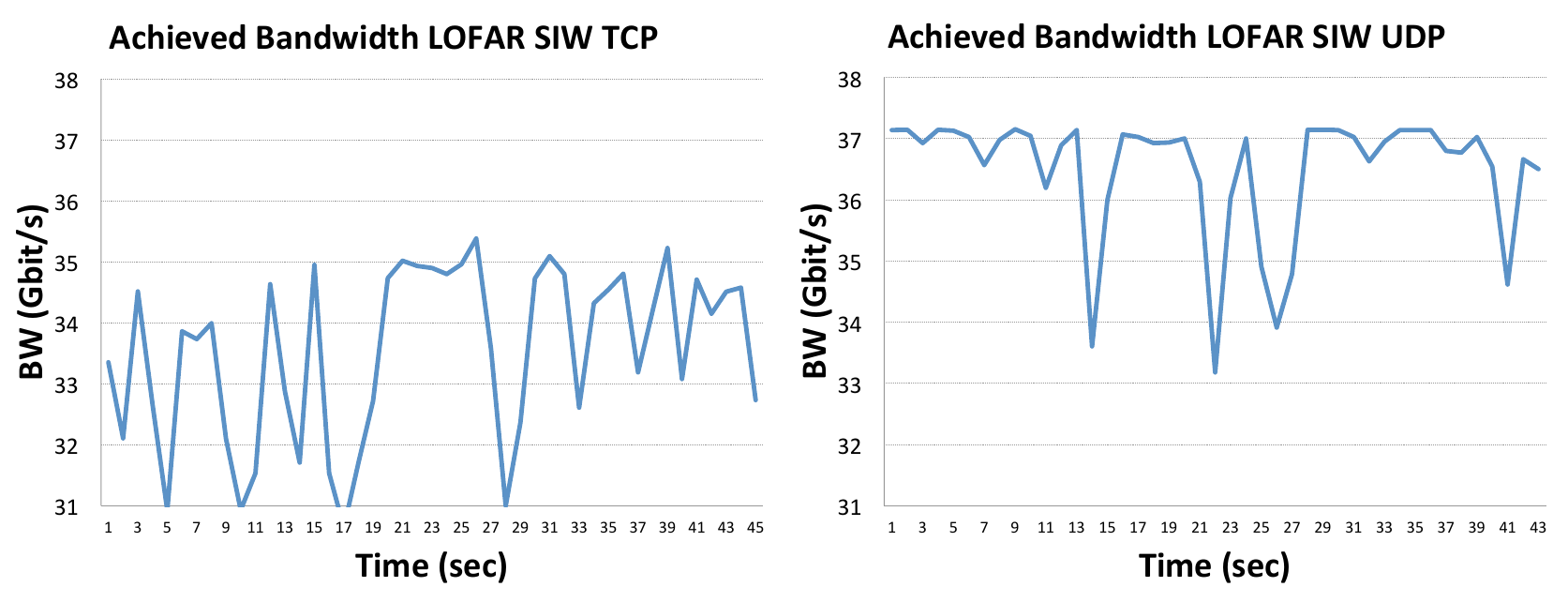}
\caption{Achieved bandwidth of LOFAR-like traffic on the receiving side using SoftiWarp TCP (left image) and SoftiWarp UDP (right image).}
\label{fig::lofarBW}
\end{figure}

In this section we mimic the data flow from LOFAR, an operational radio telescope with very similar characteristics to the future SKA.
A traffic generator is used to emulate the data produced by a LOFAR Remote Station Processing (RSP) board.
This is a UDP/IP data stream, measuring approximately 760\,Mb/s, transmitted in packets of 8\,kB, which is a limit imposed by the local memory size on the station FPGA boards. 
Each LOFAR antenna field produces four of these data streams, totalling slightly more than 3\,Gb/s per antenna field.
LOFAR currently has 78 antenna fields, 24 core stations which can be split into two independent antenna fields, and 18 remote stations, which brings the maximum LOFAR input data rate to around 250\,Gb/s.
\begin{figure}[h]
\centering
\includegraphics[width=0.95\columnwidth]{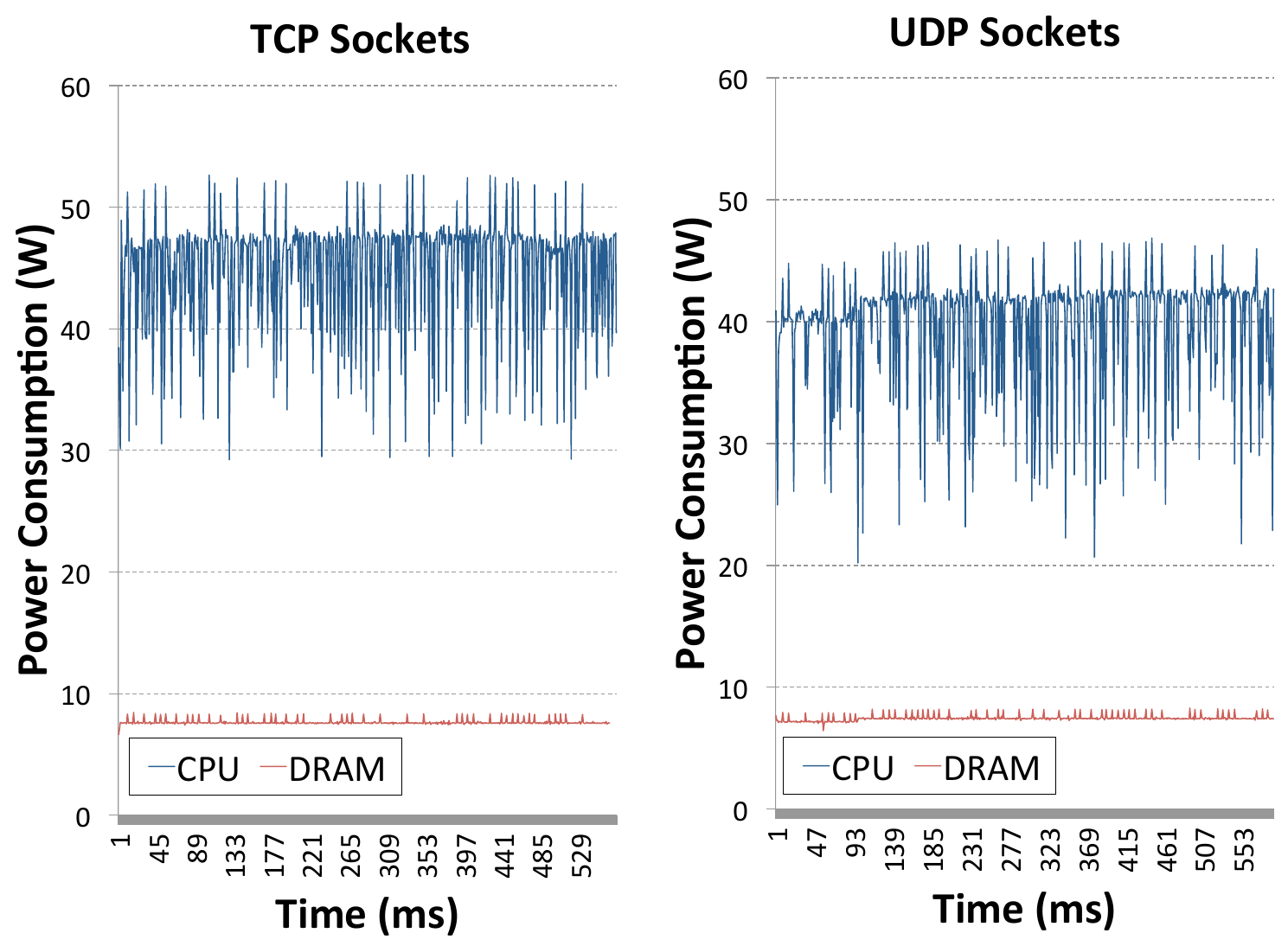}
\caption{Power consumption of CPU and DRAM for receiving a transfer of LOFAR-like traffic over TCP and UDP sockets.}
\label{fig::lofarSock}
\end{figure}
We generate 50 data streams in our experimental setup, which at 37.5\,Gb/s corresponds to roughly $\frac{1}{6}$th of the total LOFAR data flow. 
Preliminary designs of the SKA system data flow make it likely that data transported between the CSP and SDP will have very similar characteristics, albeit with much higher data rates at longer distances. Our generator is capable of transmitting the said data stream using TCP and UDP sockets and also with TCP and UDP SoftiWarp. 
\begin{figure}[h]
\centering
\includegraphics[width=\columnwidth]{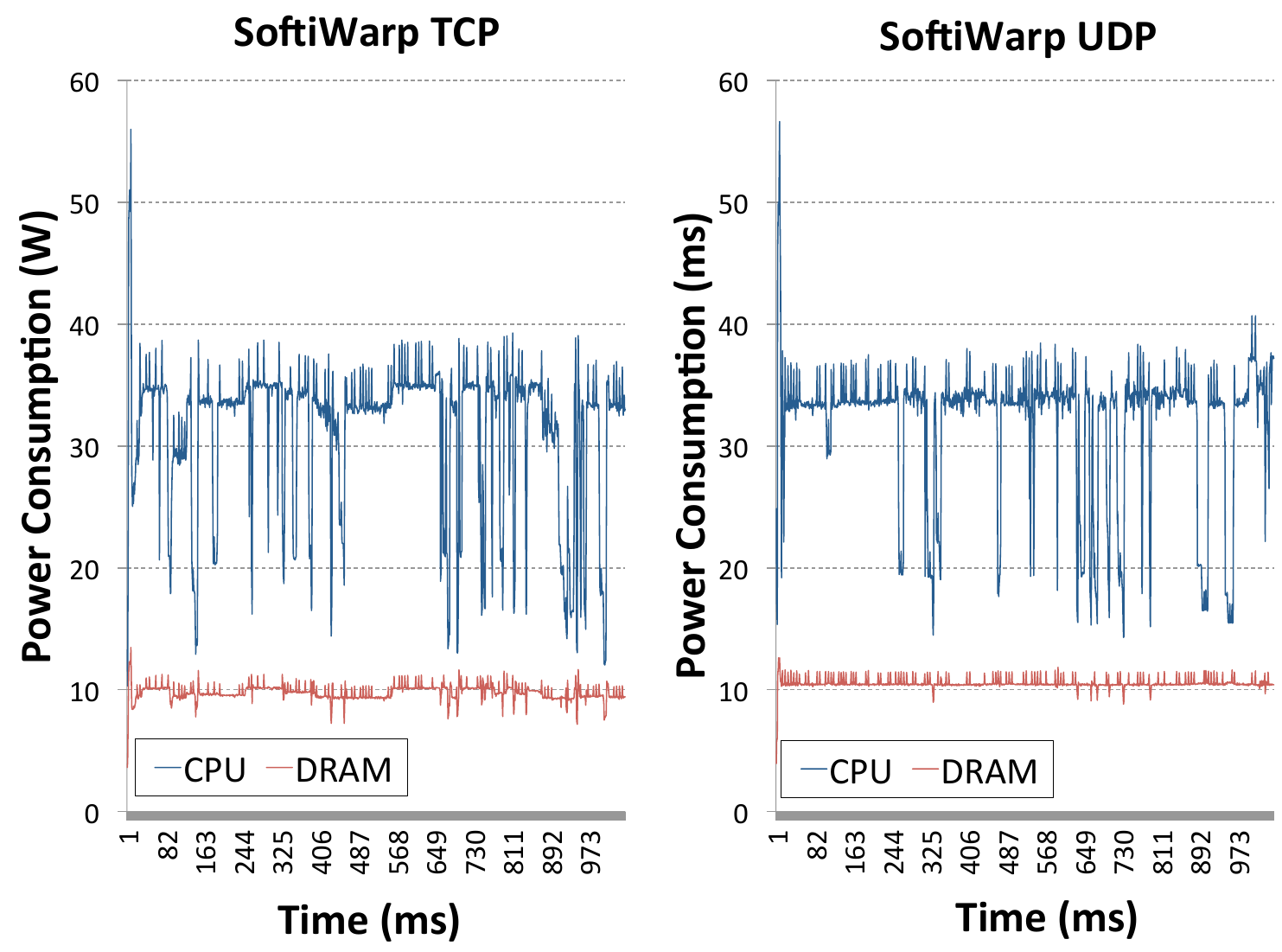}
\caption{Power consumption of CPU and DRAM for receiving a transfer of LOFAR-like traffic over SoftiWarp TCP and SoftiWarp UDP.}
\label{fig::lofarSiw}
\end{figure}
In Fig.~\ref{fig::lofarSock} we show the power consumed by receiving 50 emulated LOFAR data streams using TCP sockets in the left image, and UDP sockets in the right one. 
The energy consumed by receiving TCP traffic is measurably higher than when using UDP due to the additional overhead of the TCP/IP protocol stack.
This is a clear indication that reducing this protocol overhead will result in a smaller energy consumption.
The average power consumption of TCP in this experiment is 45.09\,W and for UDP it is 40.05\,W.
In Fig.~\ref{fig::lofarSiw} we present the power consumption measurements obtained with the LOFAR traffic generator using SoftiWarp TCP in the left image, and SoftiWarp UDP in the right image. 
The power consumption during transfers with the software iWarp implementation is clearly lower than in the case of TCP and UDP sockets, presented in Fig.~\ref{fig::lofarSock}. The average value for the TCP experiment is 32.38\,W and for the UDP experiment it is 31.01\,W.
The power efficiency difference between the TCP and UDP transfer in this case isn't as clear as with the sockets scenario, however the SoftiWarp UDP transfers achieved a better bandwidth, which can be seen in Fig.~\ref{fig::lofarBW}. We can explain this by the fact that the used message size in these transfers is relatively low (8\,kB) and the TCP-based protocol may have a problem achieving full link speed. The UDP-based protocol is more likely to achieve better speeds with smaller messages due to the lower overhead of the unreliable protocol. We will look further into the matter of achieved bandwidth in the following sections and present more results on this subject. 
\subsection{Power consumption of SoftiWARP TCP}
\label{sec:powerSoftiwarpTCP}
In this section we describe a set of transfers with the Netperf tool for the Sockets- and RDMA-based protocols using varying message sizes. This will allow us to observe the behaviour of different transport methods in different scenarios and enable us to calculate the theoretical energy efficiency for all the transport methods. The tests have been performed  with all of the offloading features of
the NIC switched off, which was done for two reasons: firstly, we want to assess the direct effect of the transport protocol on the power consumption when no hardware support is available.
Secondly, the offloading features available in modern NICs offer significantly more support for TCP protocol compared to UDP protocol, which means that with the offloading turned on the solutions based on the UDP protocol would be penalised. 
First we present the power consumption traces of different protocols and in Sec.~\ref{sec:powerEfficiency} we present the complete set of numerical values and evaluate the normalised power consumption per achieved bandwidth. 
\begin{figure}[h]
\centering
\includegraphics[width=0.95\columnwidth]{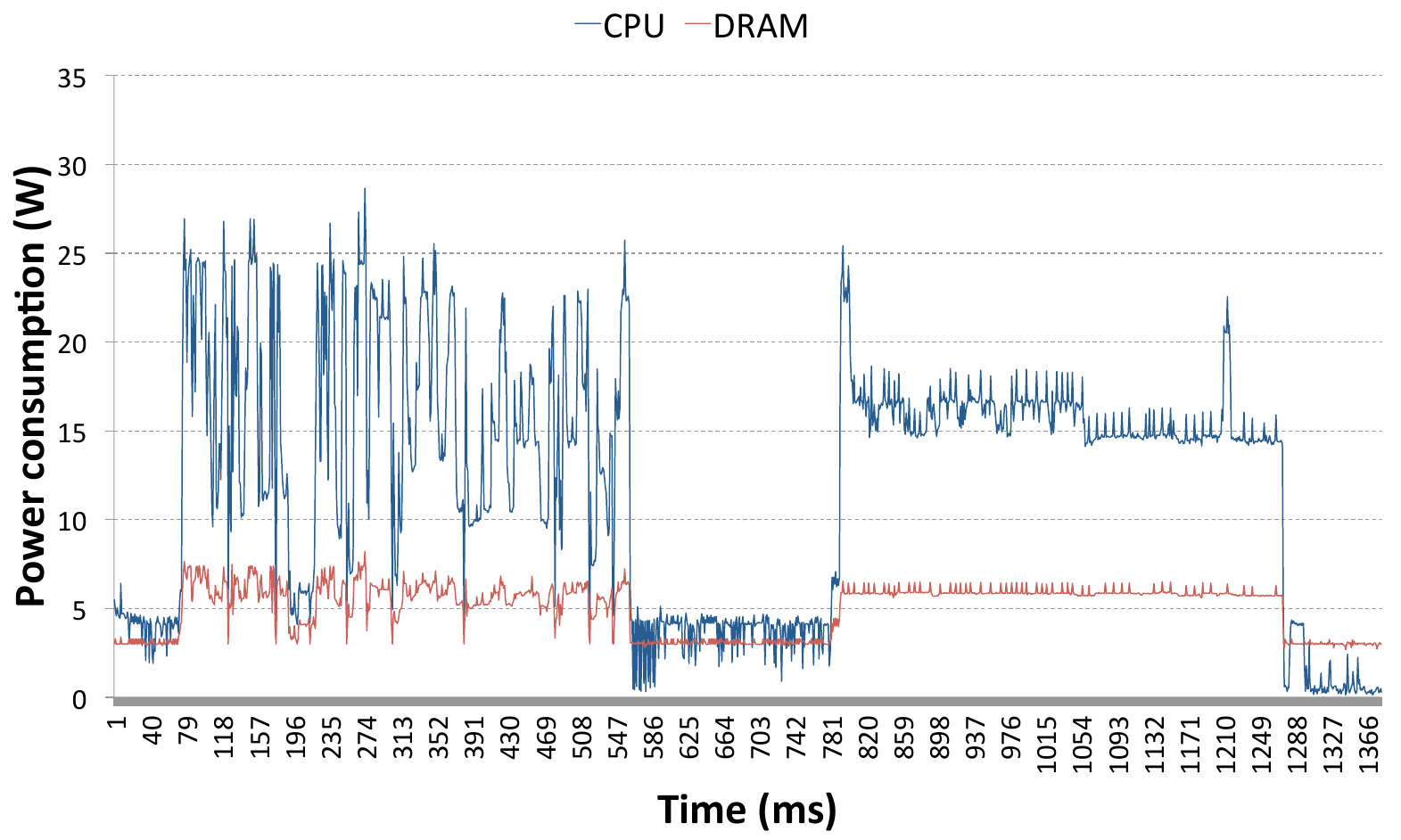}
\caption{Power consumption of CPU and DRAM for data transfer over TCP sockets
  (first peak) and SoftiWARP TCP (second peak), receiving
  side. }
\label{fig:TCPthenSiwTCPrecv}
\end{figure}
As mentioned before, we are interested in the power consumption on the
receiving side of the connection, therefore we initially focus on these results. This is motivated by the fact that the receiving sides of the data transfers in the SKA (CSP and SDP) will most likely be HPC systems, so experiments such as ours can give a good indication on the expected power consumption. Most of the sending side devices, on the other hand, will be custom-built devices. Therefore their power consumption patterns will be significantly different from a standard HPC system and the problem of their power efficiency needs to be addressed at their design level. 

Fig.~\ref{fig:TCPthenSiwTCPrecv} shows the system power trace on the receiving side during data transfer with TCP sockets (first peak) and then SoftiWARP TCP (second peak).
The blue line represents the power consumption of the CPU whereas the red line shows the DRAM power consumption.
We performed six tests for both TCP sockets and SoftiWARP TCP and compared them to confirm that the power consumption follows very similar patterns in all cases. 
The data bandwidth achieved during the tests shown in Fig.~\ref{fig:TCPthenSiwTCPrecv} is 25.1\,Gb/s for TCP sockets and 27.85\,Gb/s for SIW TCP. As we can see, neither protocol is able to achieve the full link speed when the  offloading features are switched off and the communication takes place just between two instances of the testing application. 
However, we can already note that the bandwidth achieved when using SoftiWARP TCP is slightly higher and the power consumption is lower. The average CPU power consumption from six TCP socket tests is 17.4\,W and for SoftiWARP the average is 15.89\,W.
\subsection{Power consumption of SoftiWARP UDP}
\label{sec:powerSoftiwarpUDP}
\begin{figure}[h]
\centering
\includegraphics[width=0.95\columnwidth]{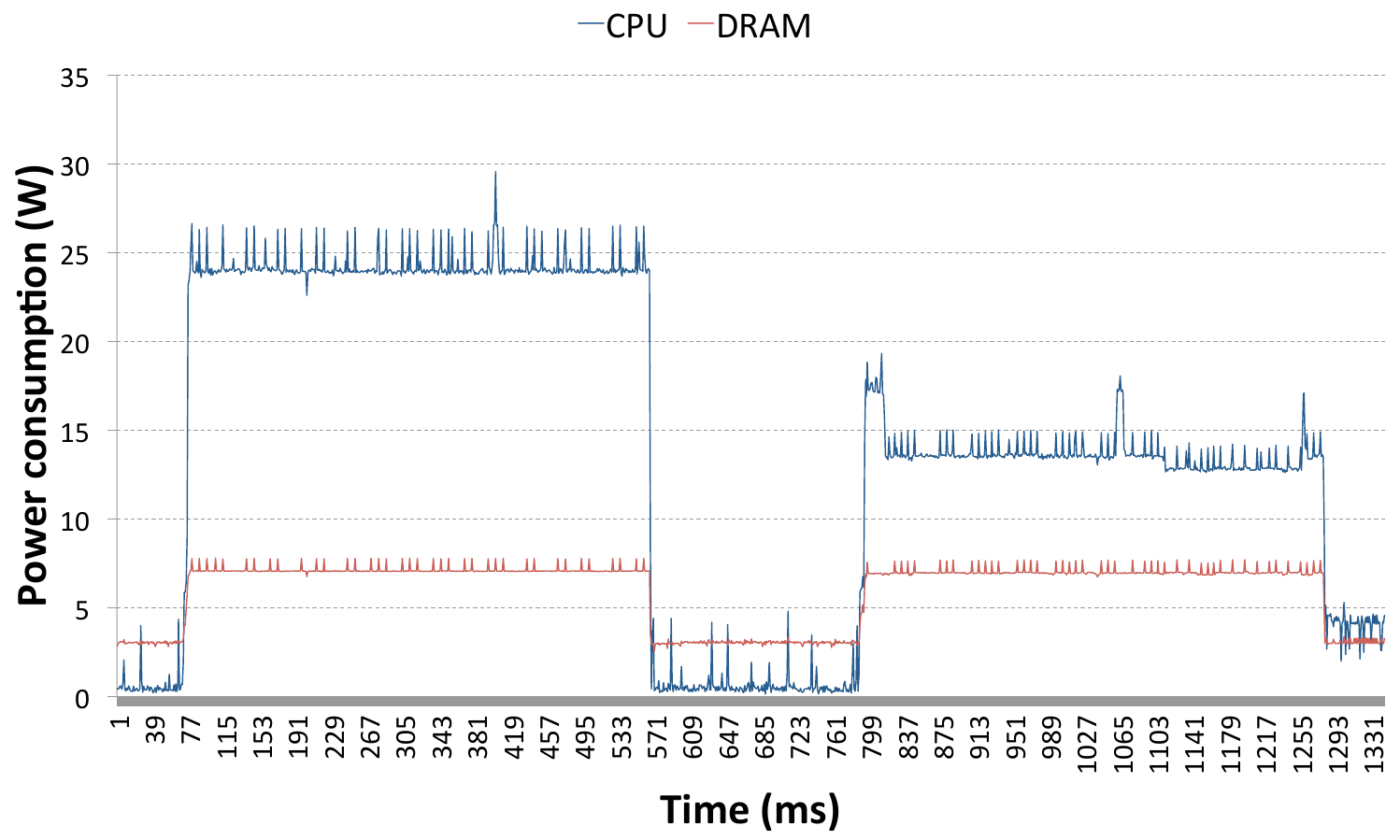}
\caption{Power consumption of CPU and DRAM for data transfer over UDP sockets
  (first peak) and SoftiWARP UDP (second peak), receiving side.}
\label{fig:UDPthenSiwUDPrecv}
\end{figure}
Similarly to Sec.~\ref{sec:powerSoftiwarpTCP} we have performed the comparison between
UDP sockets and SoftiWARP UDP. 
Fig.~\ref{fig:UDPthenSiwUDPrecv} presents the system power trace during the
execution of two Netperf tests: the first peak shows the test using UDP sockets and the second one presents a SoftiWARP UDP test. 
It is clearly visible that in this case the power consumption difference
between standard sockets and SoftiWARP is significant. The average CPU energy
consumption in the UDP sockets-based tests is 24.21\,W and 13.72\,W for the SoftiWARP-based tests. 
Furthermore, the near-full link speed of the connection is achieved in both
cases: 39.37\,Gb/s for the UDP stream test and 38.24\,Gb/s for SoftiWARP. 
\subsection{Comparison of power efficiency}
\label{sec:powerEfficiency}
In order to quantify and directly compare the power efficiency of different transport protocols we performed a set of experiments in which we measured the power consumption used by the entire data transfer, including the CPU, DRAM and the NIC. Then we calculated the normalised power efficiency, which we defined as follows:
\begin{equation} \label{eq:efficiency}
E = \frac{BW}{P}
\end{equation}
\begin{equation} \label{eq:efficiencyUnit}
[E] = \frac{\mathrm{Gb/s}}{\mathrm{W}}
\end{equation}
From equation \eqref{eq:efficiency} it can be seen that our metric, the normalised power
efficiency ($E$), is defined as the data
bandwidth ($BW$) divided by the total power consumption ($P$). From equation \eqref{eq:efficiencyUnit}
we can see that it is expressed in Gigabits per second per Watt.
With this metric we are able to provide a good comparison of how much power is needed by specific transport protocols in a manner that is independent from the variations in bandwidth in different experiments.
\begin{figure}[h]
\centering
\includegraphics[width=0.95\columnwidth]{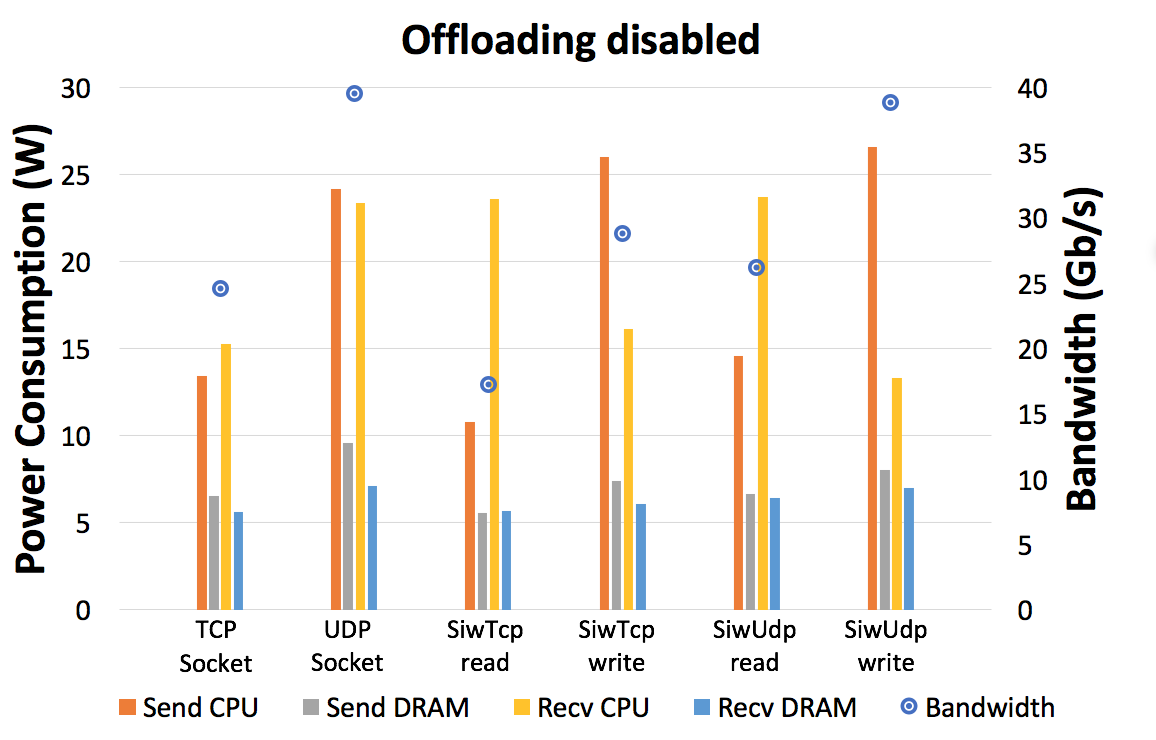}
\caption{Results of power consumption tests with the offloading features of
the NIC disabled.}
\label{fig:noOffload}
\end{figure}
\begin{figure}[h]
\centering
\includegraphics[width=0.95\columnwidth]{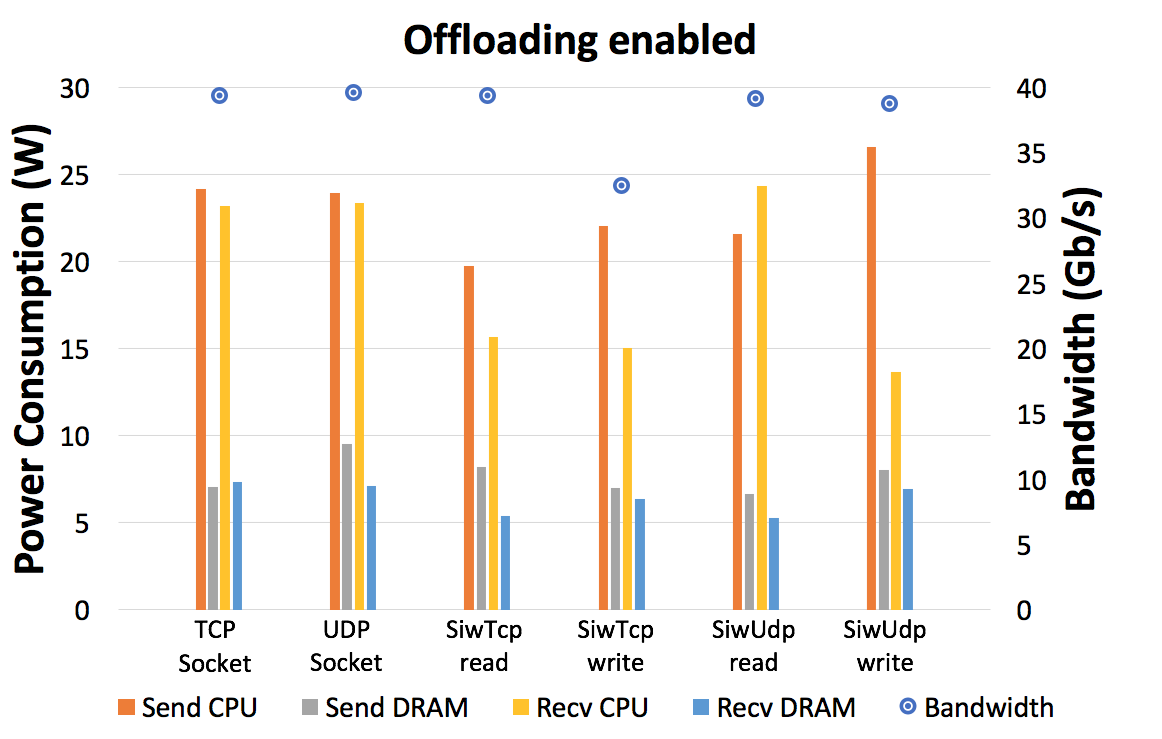}
\caption{Results of power consumption tests with the offloading features of
the NIC enabled.}
\label{fig:yesOffload}
\end{figure}
We performed six experiments for each value, using message sizes in the range of 8\,kB to
  2\,MB.
The tested transport services included: TCP sockets, UDP sockets, SoftiWARP TCP
and SoftiWARP UDP -- both using RDMA Read and RDMA Write operations. The UDP
sockets were only tested for message sizes up to 64\,kB as this size is the largest supported by this transport
protocol. As before, during the first tests all of the offloading features of the NICs were turned off. 
However, this time we also performed tests with the following offloading features enabled: rx and tx checksumming offloading, generic receive offload (GRO) and generic segmentation offload (GSO). 
This was done to see the impact of such features on the results and compare them with the no-offload scenario.

Figures~\ref{fig:noOffload}~and~\ref{fig:yesOffload} show
example results with hardware offloading features disabled and enabled,
respectively. Both figures present results for the following message sizes: 256\,kB
for TCP-based protocols and 64\,kB for the UDP-based protocols.
At these values the given protocols have achieved their maximum bandwidth.  

The tests performed without hardware offloading demonstrate that when using TCP,
even a relatively modern system is unable to achieve full link speed using a 
single core. Only the UDP-based protocols have been able to achieve the near-full
link speed; however with UDP sockets this was coupled with significant power
consumption on the sending and receiving sides. On the other hand, the SIW UDP
tests using RDMA Write have been able to achieve nearly identical results
with the hardware offloading enabled and disabled, which was around
38.79\,Gb/s bandwidth with only 13.64\,W of average power consumption on the
receiving end. The power consumption on the sending side remains among the 
highest in the above table,
but this is not a crucial issue for radio astronomy applications as the
sending side will most likely not be a standard computer but rather a
custom-built FPGA unit, designed specifically to issue RDMA Write
operations. Therefore, the power consumption of the sending side is a research
topic on its own and cannot be evaluated using experiments similar to those
presented in this paper.

The results presented in
Fig.~\ref{fig:yesOffload} confirm our assumptions from
Sec.~\ref{sec:powerSoftiwarpTCP}, namely that the TCP-based protocol family receives
significantly more support from hardware offloading. 
In the second set of tests, almost all
protocols achieved full link bandwidth, except for SIW TCP RDMA Write. The
plain UDP Socket test didn't receive any support from the hardware offloading
features, achieving the same bandwidth and power consumption. The SIW UDP RDMA
Read test achieved the full link speed due to the Receive Offload and
Segmentation Offload features. Finally, it is 
important to note that the SIW UDP Write test still offers the lowest power consumption on the receiving side of all the protocols, even when competing with the hardware-supported TCP sockets or SIW TCP. 
\begin{figure}[h]
\centering
\includegraphics[width=\columnwidth]{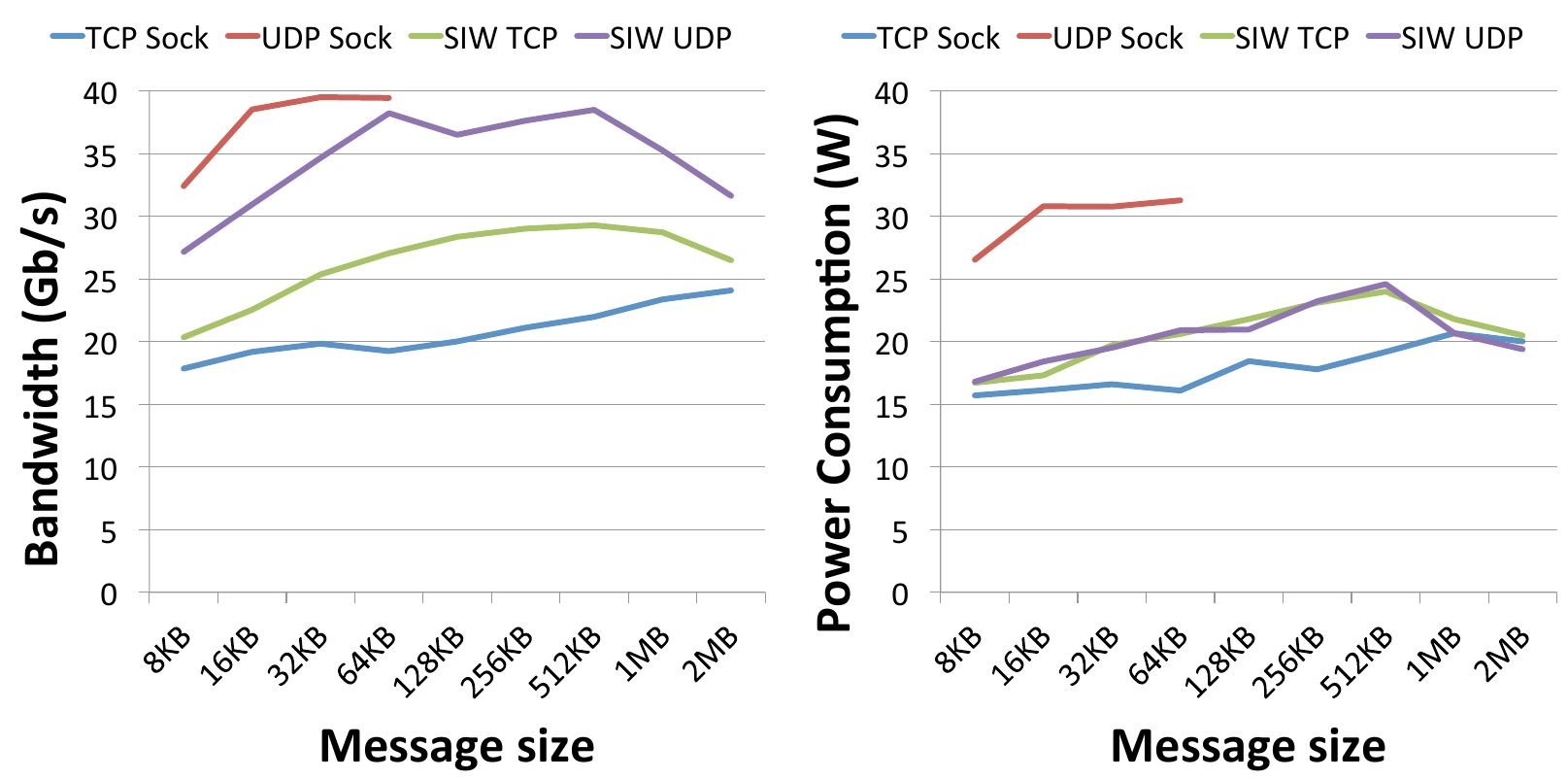}
\caption{Bandwidth (left panel) and power consumption
  (right panel) results for tests with varying message sizes.}
\label{fig:2Charts}
\end{figure}

Fig.~\ref{fig:2Charts} depicts how individual bandwidth values (left panel)
and power consumption (right panel)
correspond to varying message sizes. These charts allow us to
visualise the trends and optimal values for specific
protocols. The achieved bandwidth, but also the resulting power consumption,
increases with increasing message size for all protocols. The optimal value is
around 512\,kB for the TCP protocols and
around 64\,kB for the UDP protocols. UDP
sockets are the highest consumer of energy of all protocols, but they are also
the only ones that achieve the best link
speed without hardware support. SIW UDP is able to achieve very similar
results with regard to bandwidth, but shows much lower power consumption,
therefore its achieved power efficiency is higher.

\begin{figure}[h]
\centering
\includegraphics[width=0.95\columnwidth]{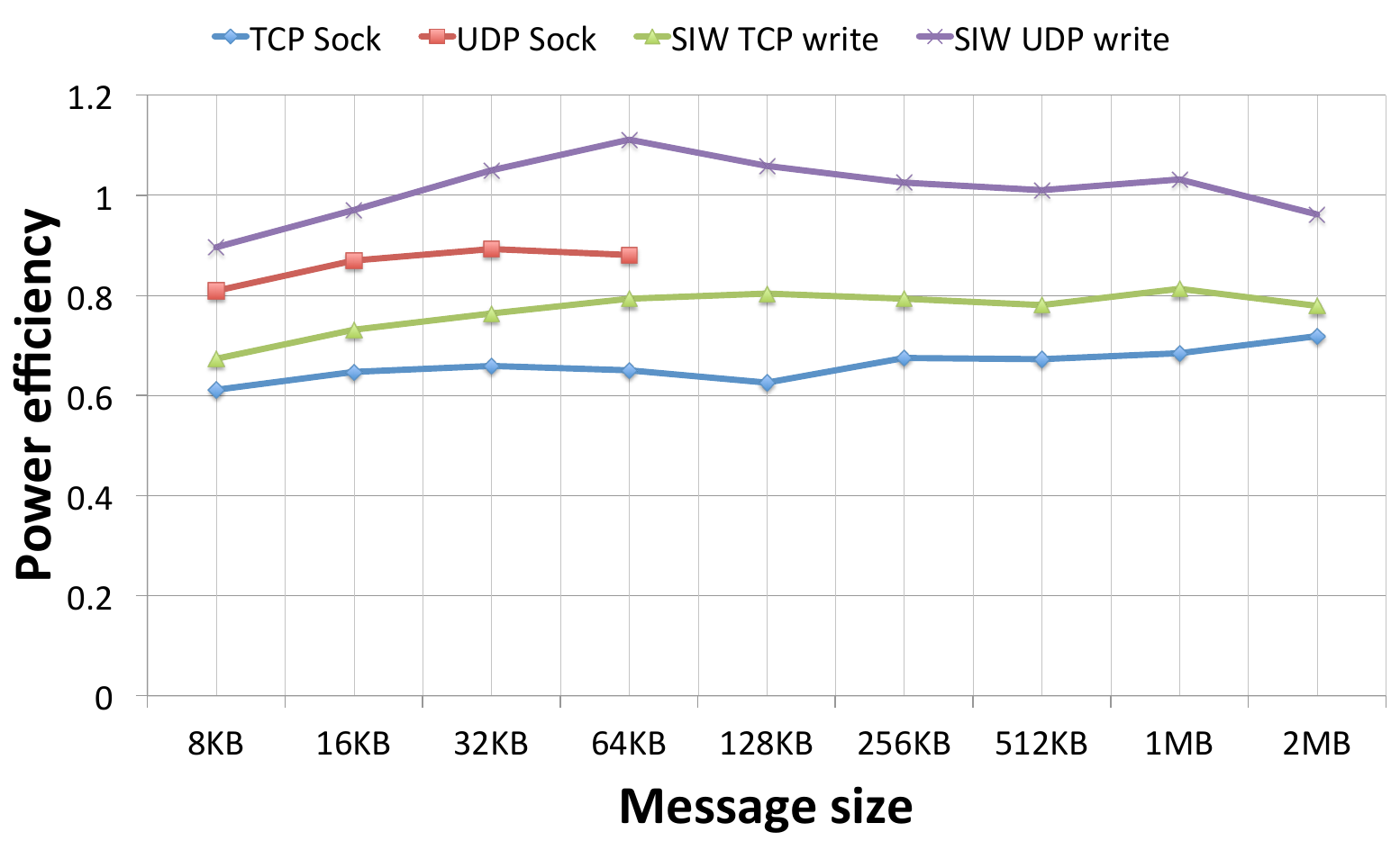}
\caption{Power efficiency (in Gigabits per second per Watt) of 10\,s Netperf
  tests, receiving side. X axis - message size, Y axis - power efficiency.}
\label{fig:powerEfficiency}
\end{figure}

The above results are used to calculate the normalised values of the power
efficiency as expressed by \eqref{eq:efficiency} and
\eqref{eq:efficiencyUnit}. The calculated values are depicted in the
efficiency chart shown in Fig.~\ref{fig:powerEfficiency}.
Comparing the TCP and UDP groups, the former one
  is less efficient, which can be explained by the low bandwidth achieved by TCP protocols as shown in Fig.~\ref{fig:2Charts} left. 
Comparing SoftiWARP protocols to plain sockets, both over TCP and UDP, we can see that SIW is more power efficient in both cases. 
In all of the experiments SIW TCP performs better than TCP sockets and SIW UDP better than UDP sockets. 
This advantage results from the design of the SoftiWARP receive path implementation: after receiving iWARP packets into the kernel memory, SoftiWARP directly copies their content into the target application buffers. Making use of the one-sided semantics of RDMA communication this final data placement does not
involve the scheduling of the receiving side application process.

Although these results are based on the software prototype of SoftiWARP UDP, we can already confirm that the reduced data touching and the decreased overhead from the OS lead to very desired characteristics and promising results. The power consumption of SoftiWARP is lower than TCP or UDP sockets in all cases and the achieved bandwidth is at least as good.
\begin{figure}[!h]
\centering
\includegraphics[width=0.95\columnwidth]{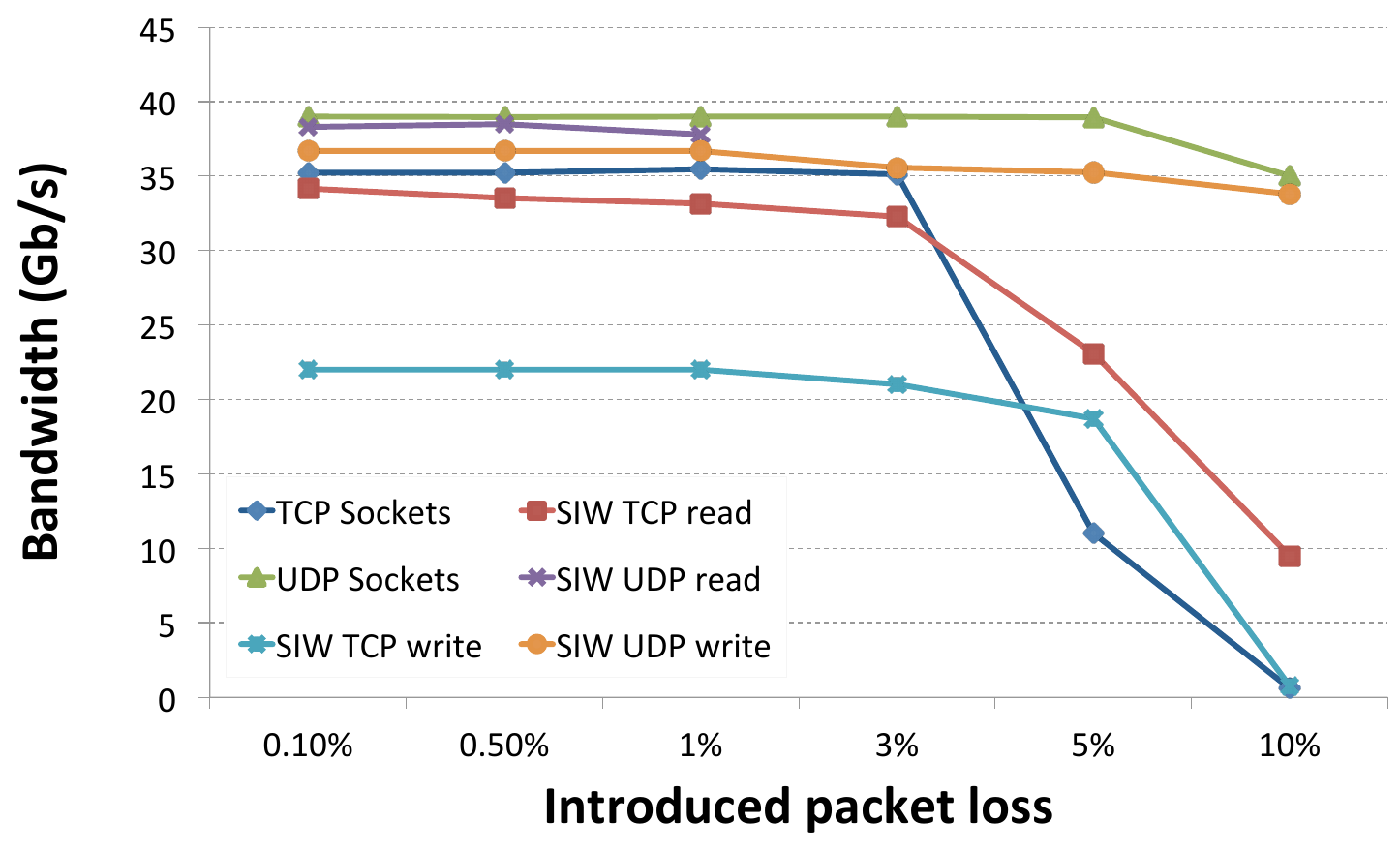}
\caption{Achieved bandwidth with introduced packet loss.}
\label{fig:packetLoss}
\end{figure}
\subsection{Behaviour in case of packet loss}
\label{sec:packetLoss}
Finally, we wanted to assess the behaviour of all the tested protocols in case
of significant packet loss. We did this by emulating packet loss using
the Netem network emulation tool\footnote{https://wiki.linuxfoundation.org/networking/netem} in the range of 0.1\% to 10\%. 
The bandwidths achieved can be seen in Fig.~\ref{fig:packetLoss}. The difference between TCP-based and UDP-based protocols is significant. The former tend to sustain their original bandwidth in the initial part of the tests as all the lost packets are re-transmitted. However, with larger packet loss the network is no longer capable of keeping up with re-transmission and the bandwidth gets significantly reduced. The UDP-based protocols do not rely on the retransmission-based reliable communication implemented by the TCP protocol and are able to maintain the transfers on the same level, regardless of the problems occurring along the link. The only decrease in bandwidth is the actual amount of packets that have been dropped. As we can see in the chart, this doesn't hold true for the results of SIW UDP RDMA Reads, which - as discussed earlier - are not yet fully supported in our implementation. The current protocol does not recover from completely lost RDMA READ request/response pairs, which results in transfer breakdown as soon as the packet loss reaches 3\%.

The above results show that the use of a protocol that relies on two-way communication and tries to provide full reliability on the transport level, such as the TCP, can be infeasible for a scenario such as the SKA. 
It is true that the introduced packet loss in our experiments was very high, but the tests were performed for a short, local connection. 
In the case of the SKA, where the connections spread over hundreds of kilometres in length, we would see a much more drastic influence of packet loss on the achieved bandwidth. 
This result confirms another reason for the choice of an unreliable transport protocol for our purposes. 
The power consumption in different packet loss scenarios didn't show any
noteworthy behaviour. 
It corresponded to what we have seen in our previous experiments, namely that with growing packet loss the energy consumption was lower, because the achieved bandwidth was also lower.

\section{Conclusions and Future Work}
In this paper we presented the data transport requirements of the world's largest
radio telescope, the Square Kilometre Array (SKA).
We proposed a solution to meet these requirements, namely an unreliable, datagram-based iWARP
protocol implementation. We then presented a software prototype of such a
protocol, called SoftiWARP UDP, and evaluated its performance and power
efficiency together with those of TCP and UDP sockets. 
We have confirmed that UDP is a very good choice for long distance transfer of astronomical data. The protocol overhead is lower, which leads to lower power consumption. Furthermore, the use of a reliable transport protocol is not feasible in a scenario such as the SKA, as it (1) leads to higher power consumption, and (2) the data transfer quality soon becomes unacceptable in the case of non-negligible data packet loss.

Our software prototype of SoftiWARP UDP is already capable of outperforming TCP and UDP sockets in terms of power efficiency. This is a very desired result, however we expect a much higher improvement of the power efficiency with implementation of the SoftiWARP UDP protocol in hardware, e.g. using FPGAs and the source code of SoftiWARP UDP, which we leave for future work on this subject. A desired solution for the SKA purposes would have typical RDMA characteristics, where all four lower network layers are handled in hardware. Considering our results, we believe that with a hardware implementation the SoftiWARP UDP protocol would bring all the benefits of RDMA, namely an outstanding power efficiency, low latencies and CPU utilization and high bandwidths, while meeting the specific requirements of the radio astronomy data transfer service. 

For the future work on this topic we are planning to look into
using flash storage technology for data ingress, which is
energy efficient and offers high bandwidth and low-latency access.
\section*{Acknowledgment}
This work is conducted in the context of the joint ASTRON and IBM DOME project 
and is funded by the Netherlands Organisation for Scientific Research (NWO), 
the Dutch Ministry of Economic Affairs (EL\&I), and the Province of Drenthe.

\bibliography{references}

\hfill \break
\hfill \break
\text{*} Linux is a registered trademark of Linus Torvalds in the United States, other countries, or both.
Intel and Intel Xeon are trademarks or registered trademarks of Intel Corporation or its subsidiaries in
the United States or other countries.
IBM and IBM BlueGene/P are trademarks of International Business Machines Corporation, registered in many
jurisdictions worldwide. Other product or service names may be trademarks or service marks of IBM or
other companies.

\end{document}